\documentclass{PoS}

\title{The three-quark potential and perfect Abelian dominance 
in SU(3) lattice QCD}

\ShortTitle{The three-quark potential and perfect Abelian dominance 
in SU(3) lattice QCD}

\author{\speaker{Hideo Suganuma}, \\
        Department of Physics \& Division of Physics and Astronomy, 
Graduate School of Science, \\
Kyoto University, 
Kitashirakawaoiwake, Sakyo, Kyoto 606-8502, Japan\\
        E-mail: \email{suganuma@scphys.kyoto-u.ac.jp}}

\author{Naoyuki Sakumichi \\
Ochanomizu University, 
2-1-1 Otsuka, Bunkyo, Tokyo 112-8610, Japan}

\abstract{
We study the static three-quark (3Q) potential 
for more than 300 different patterns of 3Q systems with 
high statistics, i.e., 1000-2000 gauge configurations, 
in SU(3) lattice QCD at the quenched level. 
For all the distances, the 3Q potential is found to be well described by 
the Y-ansatz, i.e., one-gluon-exchange (OGE) Coulomb 
plus Y-type linear potential. 
Also, we investigate Abelian projection of quark confinement 
in the context of the dual superconductor picture proposed by 
Yoichiro~Nambu~{\it et al.} in SU(3) lattice QCD. 
Remarkably, quark confinement forces in both Q$\bar{\rm Q}$ 
and 3Q systems can be described only with Abelian variables 
in the maximally Abelian gauge, i.e., 
$\sigma_{\rm Q \bar Q} \simeq \sigma_{\rm Q \bar Q}^{\rm Abel} \simeq \sigma_{\rm 3Q} \simeq 
\sigma_{\rm 3Q}^{\rm Abel}$, 
which we call ``perfect Abelian dominance'' of quark confinement.
}

\FullConference{The 33rd International Symposium on Lattice Field Theory\\
                 14 -18 July  2015\\
                 Kobe International Conference Center, Kobe, Japan}

\begin{document}

\section{Introduction}

In 1966, Yoichiro Nambu \cite{N66} first proposed the SU(3) gauge theory, 
i.e., quantum chromodynamics (QCD), as a field theory of quarks, 
just after the introduction of the color quantum number \cite{HN65}.
In 1973, the asymptotic freedom of QCD was theoretically shown \cite{GWP73}, 
and QCD was established as the fundamental theory of the strong interaction.
While perturbative QCD works well at high energies, 
infrared QCD exhibits strong-coupling nature and 
various nonperturbative phenomena 
such as dynamical chiral-symmetry breaking \cite{NJL61} and 
color confinement \cite{N74}. 

Among the nonperturbative properties of QCD, color confinement is 
one of the most difficult important subjects. 
The difficulty is considered to originate from non-Abelian dynamics 
and nonperturbative features of QCD, which are largely different from QED. 
However, it is not clear whether quark confinement is peculiar to 
the non-Abelian nature of QCD or not.

On the quark confinement in hadrons, 
Q$\bar{\rm Q}$ systems have been well investigated 
in lattice QCD \cite{R12}, but 
the quark interaction in baryonic three-quark (3Q) systems 
\cite{TS0102, TS0304} has not been studied so much. 
Note however that the nucleon is one of the main ingredients of 
the matter in our real world, 
and therefore the quark confinement in baryons would be fairly important.
Furthermore, the three-body force among three quarks is a ``primary'' force 
reflecting the SU(3) gauge symmetry in QCD \cite{TS0102}, 
while the three-body force appears as a residual interaction 
in most fields of physics. 

In this paper, we accurately measure the static 3Q potential 
and quark confinement in baryons in SU(3) quenched lattice QCD with
1000-2000 gauge configurations \cite{SS15}. 
In parallel, we also investigate Abelian projection 
of quark confinement 
for both Q$\bar{\rm Q}$ and 3Q systems \cite{SS15,SS14}.

\section{Dual Superconductor Picture and Maximally Abelian projection}

In 1970's, Nambu, 't~Hooft and Mandelstam proposed 
a dual-superconductor theory for quark confinement \cite{N74}. 
In this theory, the QCD vacuum is identified as 
a color-magnetic monopole condensed system, and 
there occurs one-dimensional squeezing of the color-electric flux 
among (anti)quarks by the dual Meissner effect, 
which leads to the string picture \cite{N6970} of hadrons.

However, there are two large gaps between QCD and the 
dual-superconductor picture \cite{IS99}.
\begin{enumerate}
\item
The dual-superconductor picture is based on the Abelian gauge theory 
subject to the Maxwell-type equations, 
while QCD is a non-Abelian gauge theory.
\item
The dual-superconductor picture requires 
color-magnetic monopole condensation as the key concept, 
while QCD does not have such a monopole as the elementary degrees of freedom.
\end{enumerate}
As a connection between the dual superconductor and QCD, 
't~Hooft proposed ``Abelian projection'' \cite{tH81,EI82}, 
which accompanies topological appearance of magnetic monopoles. 
't~Hooft also conjectured that long-distance physics such as confinement 
is realized only by Abelian degrees of freedom in QCD \cite{tH81}, 
which is called ``(infrared) Abelian dominance''. 
Actually, in the maximally Abelian (MA) gauge \cite{KSW87}, 
infrared QCD becomes Abelian-like \cite{SY90}
as a result of a large off-diagonal gluon mass of about 1GeV \cite{AS99}, 
and also there appears a large clustering of 
the monopole-current network in the QCD vacuum \cite{KSW87,SNW94}.
In fact, by taking the MA gauge, infrared QCD seems to resemble 
an Abelian dual-superconductor system. 
In the MA gauge, Abelian dominance of quark confinement 
has been investigated mainly for Q$\bar{\rm Q}$ systems 
in SU(2) and SU(3) lattice QCD \cite{SY90,STW02,DIK04}.

Lattice QCD is described with 
the link variable $U_\mu (s)=e^{iagA_\mu(s)}$ 
($a$:~lattice spacing, $g$:~gauge coupling, $A_\mu$:~gluon fields), 
and SU(3) MA gauge fixing \cite{SS15,SS14} 
is performed by maximizing
\vspace{-0.1cm}
\begin{eqnarray}
 R_{\rm MA}[U_\mu(s)]
\equiv \sum_{s} \sum_{\mu=1}^4  
{\rm tr}\left( U_\mu^\dagger(s)\vec H U_\mu(s)\vec H\right)  
= \frac{1}{2} \sum_{s} \sum_{\mu=1}^4  
\left(  \sum_{i=1}^3 |U_\mu(s)_{ii} |^2 -1 \right), 
\label{MAgf}
\end{eqnarray}
under the SU(3) gauge transformation. 
In our calculation, we numerically maximize $R_{\rm MA}$ 
using the over-relaxation method \cite{SS15,SS14,STW02}. 
The converged value of 
$\langle R_{\rm MA} \rangle/(4V)\in [-\frac{1}{2},1]$ 
($V$: lattice volume) is, e.g., 
$0.7072(6)$ at $\beta=5.8$ and $0.7322(5)$ at $\beta=6.0$, 
and the maximized value of $R_{\rm MA}$ is almost the same over 
1000-2000 gauge configurations. 
Then, our procedure seems to escape bad local minima, 
where $R_{\rm MA}$ is relatively small, 
so that the Gribov copy effect would not be significant.

The Abelian link variable 
$u_\mu(s) = 
e^{i\theta_\mu^3(s) T_3 + i\theta_\mu^8(s) T_8}
\in {\rm U(1)}^2$
is extracted from the link variable
$U_\mu^{\rm MA} (s) \in$ SU(3) in the MA gauge, 
by maximizing 
$R_{\rm Abel} \equiv \frac{1}{3}
{\rm Re} \, {\rm tr}\left( U_\mu^{\rm MA}(s) u_\mu^\dagger(s) \right) 
\in [-\frac{1}{2},1]$ \cite{SS15}.

\section{The quark-antiquark potential 
and perfect Abelian dominance of confinement}

First, we investigate the Q$\bar{\rm Q}$ potential $V(r)$ 
in SU(3) quenched lattice QCD on $L^3 \times L_t$, with 
$(\beta, L^3 L_t)=(6.4, 32^4), (6.0, 32^4)$ and $(5.8, 16^3 32)$ \cite{SS14}.
The static Q$\bar{\rm Q}$ potential $V(r)$ 
is obtained from the Wilson loop 
\cite{R12}, and its MA projection (Abelian part) $V_{\rm Abel}(r)$ is similarly defined as 
\vspace{-0.1cm}
\begin{eqnarray}
V(r)=-\lim_{t \rightarrow \infty} 
\frac{1}{t}{\rm ln} \langle W[U_\mu] \rangle_{r \times t}, \qquad
V_{\rm Abel}(r)=-\lim_{t \rightarrow \infty} \frac{1}{t}{\rm ln} 
\langle W[u_\mu] \rangle_{r \times t}.
\end{eqnarray}
(We also define the off-diagonal part $V_{\rm off}(r)$, 
and numerically find 
$V(r) \simeq V_{\rm Abel}(r) + V_{\rm off}(r)$ \cite{SS14}.)

We show in Fig.~1 the lattice QCD result of 
the Q$\bar{\rm Q}$ potential $V(r)$ 
and its Abelian part $V_{\rm Abel}(r)$. They are 
found to be well reproduced by the Coulomb-plus-linear ansatz, respectively:
\vspace{-0.1cm}
\begin{eqnarray} 
V(r) = - \frac{A}{r} + \sigma r + C,  
\qquad
V_{\rm Abel}(r) 
= - \frac{A_{\rm Abel}}{r} + \sigma_{\rm Abel}~r + C_{\rm Abel}. 
\label{eq:Cornell}
\end{eqnarray}
Remarkably, we find ``perfect Abelian dominance'' 
of the string tension, $\sigma_{\rm Abel} \simeq \sigma$, on these lattices. 

%%%%%%% FIGURE: Main results %%%%%%%%%%%%%%%%%%%%%%%
\begin{figure}[hb]
\centering
\includegraphics[width=13cm,clip]{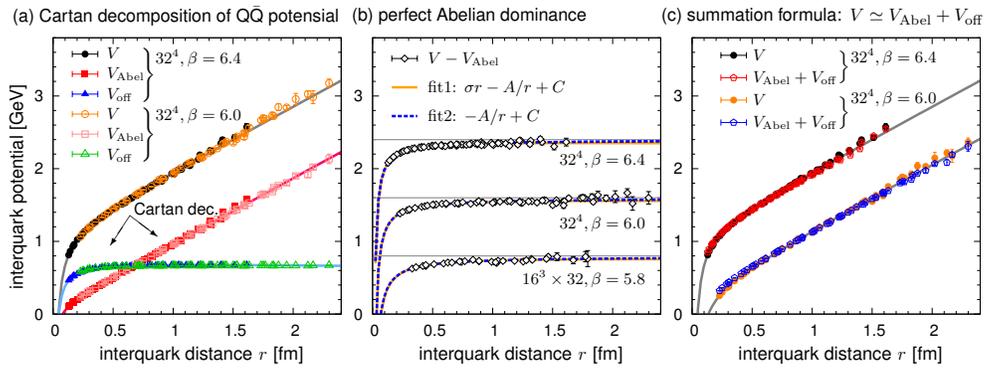}
\caption{(a) Cartan decomposition of the static Q$\bar{\rm Q}$ potential 
$V(r)$ (circles) into the Abelian part $V_{\rm Abel}(r)$ (squares) 
and the off-diagonal part $V_{\rm off}(r)$ (triangles) 
on $32^4$ at $\beta=6.4$ (filled) and $6.0$ (open). 
For each potential, 
the best-fit Coulomb-plus-linear curve is added. 
(b) Fit analysis of $V(r) - V_{\rm Abel}(r)$ at $\beta$=6.4, 6.0 and 5.8. 
At each $\beta$, all the data can be well fit with the pure Coulomb form 
with $\sigma=0$.
(c) The demonstration of 
$V(r) \simeq V_{\rm Abel}(r) + V_{\rm off}(r)$  
at $\beta =6.0$ (upper) and $6.4$ (lower). 
All the figures are taken from Ref.~\cite{SS14}.
}
\end{figure}
%%%% End FIGURE: Main results %%%%

We also examine the physical lattice-volume dependence 
of $\sigma_{\rm Abel}/\sigma$ in Fig.~2. 
Perfect Abelian dominance ($\sigma_{\rm Abel}/\sigma \simeq 1$) seems 
to be realized when the spatial size $La$ is larger than about $2$~fm.

\begin{figure}[h]
\centering
\includegraphics[width=6cm,clip]{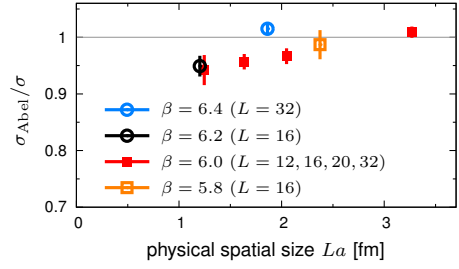}
\caption{
Physical spatial-size dependence of the ratio $\sigma_{\rm Abel}/\sigma$ 
of the Q$\bar{\rm Q}$ string tension and the Abelian one, 
taken from Ref.~\cite{SS15}. 
(For $\beta=5.8$ and $6.0$, both statistics and number of data 
in Ref.~\cite{SS15} are larger than those in Ref.~\cite{SS14}.)
Perfect Abelian dominance ($\sigma_{\rm Abel}/\sigma \simeq 1$) is found 
for larger lattices with $La \ge 2$~fm.
}
\end{figure}

\section{The baryonic three-quark potential}

In this section, we perform the accurate calculation of 
the baryonic three-quark (3Q) potential $V_{\rm 3Q}$
in SU(3) quenched lattice QCD 
with the standard plaquette action on the two lattices \cite{SS15}:

i) $\beta =5.8$ on $16^3 \times 32$,~~~ 
[i.e., $a= 0.148(2)$~fm,~ the spatial volume $(La)^3=(2.37(3)$~fm$)^3$],

ii) $\beta =6.0$ on $20^3 \times 32$,~~
 [i.e., $a=0.1022(5)$~fm, the spatial volume $(La)^3=(2.05(1)$~fm$)^3$].

\noindent
The lattice spacing $a$ is determined 
from the string tension $\sigma =0.89$ GeV/fm 
in the Q$\bar{\rm Q}$ potential.
For $\beta$=5.8 and 6.0, 
we use $2000$ and $1000$ gauge configurations, 
respectively, which are taken every $500$ sweeps 
after a thermalization of $20000$ sweeps.
The jackknife method is used for the error estimate.

\begin{figure}[h]
\centering
\includegraphics[width=8.6cm,clip]{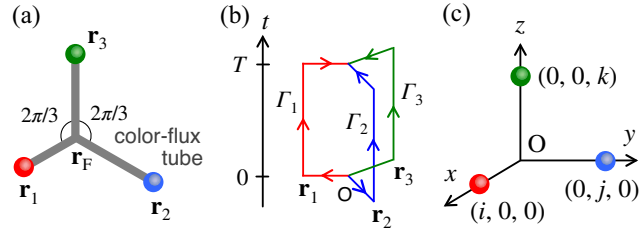}
\caption{
(a) The flux-tube configuration of the three-quark system with the minimal value of the total flux-tube length, $L_{\rm min}$. ${\bf r}_F$ is the Fermat point.
(b) The trajectory of the 3Q Wilson loop $W_{\rm 3Q}$.
(c) The location of the static three-quark sources 
in our lattice QCD calculation. These figures are taken from Ref.\cite{SS15}.}
\end{figure}

Similar to the case of the Q$\bar{\rm Q}$ potential $V(r)$, 
the color-singlet baryonic 3Q potential $V_{\rm 3Q}$ 
can be calculated from the 3Q Wilson loop $W_{\rm 3Q}$ as \cite{TS0102,TS0304} 
\begin{equation}
V_{\rm 3Q} = - \lim_{T\rightarrow \infty}\frac{1}{T}\ln \left\langle 
W_{\rm 3Q} \left[ U_\mu  \right] \right\rangle, \quad
W_{\rm 3Q} \left[ U_\mu \right]
\equiv \frac{1}{3!} \sum_{a,b,c} \sum_{a^\prime b^\prime c^\prime} \epsilon_{abc}\epsilon_{a^\prime b^\prime c^\prime}
X_1^{aa^\prime}X_2^{bb^\prime}X_3^{cc^\prime}.
\end{equation}
Here, $X_k \equiv \prod_{\Gamma_k}  U_\mu (s)$
is the path-ordered product of the link variables 
along the path $\Gamma_k$ in Fig.~3(b). 
We put three quarks on 
$(i,0,0)$, $(0,j,0)$ and $(0,0,k)$ in ${\bf R}^3$ 
with $1\leq i \leq j \leq k \leq L/2$ in lattice units, 
as shown in Fig.~3(c), 
and deal with 101 and 211 different patterns 
of 3Q systems at $\beta$=5.8 and 6.0, respectively, 
based on well-converged data of $\left\langle W_{\rm 3Q} \right\rangle$. 
For the accurate calculation of the 3Q potential 
with finite $T$, we apply 
the gauge-invariant smearing method \cite{TS0102,TS0304}, 
which enhances the ground-state component in the 3Q state in 
$\langle W_{\rm 3Q} \rangle$.

\begin{table}[h]
\caption{Fit analysis of inter-quark potentials 
in lattice units 
at $\beta =5.8$ on $16^3 \times 32$ 
and $\beta=6.0$ on $20^3 \times 32$.
$N_Q$ is the number of different patterns of Q$\bar{\rm Q}$ or 3Q systems.
For the Q$\bar{\rm Q}$ potential $V$ and the Abelian part $V^{\rm Abel}$, 
we list the best-fit parameter set $(\sigma, A)$ of 
the Coulomb-plus-linear ansatz. 
For the 3Q potential $V_{\rm 3Q}$ and 
the Abelian part $V_{\rm 3Q}^{\rm Abel}$, 
we list the best-fit parameter set 
$(\sigma_{\rm 3Q}, A_{\rm 3Q})$ of the Y-ansatz. 
``3Q(equi.)'' means the fit only for 
equilateral-triangle 3Q systems.
The string tension ratio $\sigma^{\rm Abel}/\sigma$ 
is also listed. (See Ref.[9].)
}
\begin{tabular}{cccccccc}
\hline 
$\beta$ & 
 & $N_{\rm Q}$
 &  \qquad $\sigma $  & \,\,$A$ 
 &  \quad $\sigma^{\rm Abel}$  & $A^{\rm Abel}$  
 &  $\sigma^{\rm Abel}/\sigma $  \\
\hline 
5.8 
    & QQbar & 26 
    & \,\, 0.099(2) & 0.30(3)  & \,\, 0.098(1) & 0.043(12) 
    & \,\, 0.99(3)       \\
& 3Q (equi.) & 5
    & \,\, 0.097(1) & 0.118(3)  & \,\, 0.098(3) & $-$0.001(8) 
    & \,\, 1.01(3)       \\
& 3Q & 101
    & \,\, 0.0997(4) & 0.109(1) & \,\, 0.0967(5) & 0.006(2)  
    & \,\, 0.97(1)       \\ 
\hline
6.0 
    & QQbar & 39 
    & \,\, 0.0472(6) & 0.289(10) & \,\, 0.0457(2) & 0.050(3)  
    & \,\, 0.97(1)       \\
& 3Q (equi.) & 8
    & \,\, 0.0471(10) & 0.121(3) & \,\, 0.0455(12) & 0.014(4) 
    & \,\, 0.97(3)       \\
& 3Q & 211
    & \,\, 0.0480(3) & 0.113(1) & \,\, 0.0456(2) & 0.013(1) 
    & \,\, 0.95(1)       \\
\hline 
\end{tabular}
\end{table}

As the result, 
we find that the 3Q potential $V_{\rm 3Q}$ 
is fairly well reproduced by 
the Y-ansatz \cite{TS0102,TS0304}, 
i.e., one-gluon-exchange Coulomb plus Y-type 
linear potential, 
\begin{eqnarray} 
V_{\rm 3Q} ({\bf r}_1, {\bf r}_2, {\bf r}_3)
&=& -\sum_{i<j} \frac{A_{\rm 3Q} }{|{\bf r}_i-{\bf r}_j|}
+\sigma_{\rm 3Q} L_{\rm min}+C_{\rm 3Q}
=- \frac{A_{\rm 3Q}}{R}
+\sigma_{\rm 3Q} L_{\rm min}+C_{\rm 3Q},
\label{eq:Y-ansatz}
\end{eqnarray}
for all the distances of the 3Q systems \cite{TS0102,TS0304,SS15}.
Here, ${\bf r}_1, {\bf r}_2$ and ${\bf r}_3$ denote the three-quark 
positions, and $L_{\rm min}$ is the minimum flux-tube length 
connecting the three quarks, as shown in Fig.~3(a).
Here, we have introduced a convenient variable 
$1/R \equiv \sum_{i<j} 1/|{\bf r}_i-{\bf r}_j|$.

Table~1 is a summary of the fit analysis for 
the 3Q potential $V_{\rm 3Q}$ with the Y-ansatz 
and the Q$\bar{\rm Q}$ potential $V$ 
with Eq.(\ref{eq:Cornell}) 
in SU(3) lattice QCD at $\beta=5.8$ on $16^3 \times 32$ and 
$\beta=6.0$ on $20^3 \times 32$ \cite{SS15}.

As shown in Fig.3(a), the functional form (\ref{eq:Y-ansatz}) indicates 
the Y-shaped flux-tube formation in baryons. 
Actually, the Y-shaped flux-tube formation has been observed 
in the lattice QCD calculations on the action density 
in the presence of static three quarks \cite{DIK04,IBSS03}.

\section{Perfect Abelian dominance of quark confinement in baryons}

In this section, we investigate Abelian dominance of quark confinement 
in the 3Q system.
Similarly to the Q$\bar{\rm Q}$ case, 
the MA-projected 3Q potential $V_{\rm 3Q}^{\rm Abel}$ (Abelian part)
can be calculated from the Abelian 3Q Wilson loop 
$W_{3Q} \left[u_\mu \right]$ in the MA gauge:
\begin{eqnarray}
V_{\rm 3Q}^{\rm Abel} = - \lim_{T\rightarrow \infty}
\frac{1}{T}\ln \left\langle 
W_{\rm 3Q} \left[u_\mu \right] \right\rangle, 
\end{eqnarray}
which is invariant 
under the residual Abelian gauge transformation.

Figure~4 shows the 3Q potential $V_{\rm 3Q}$ and 
its Abelian part $V_{\rm 3Q}^{\rm Abel}$ 
plotted against $L_{\rm min}$ 
in SU(3) lattice QCD at $\beta$=5.8 on $16^3 \times 32$ \cite{SS15}. 
For comparison, we show in Fig.4(a) the Q$\bar{\rm Q}$ potential $V(r)$ 
and its Abelian part $V^{\rm Abel}(r)$, indicating 
perfect Abelian dominance of the string tension in mesons.

We note that the Abelian dominance of the Q$\bar{\rm Q}$ confinement force 
does not necessarily mean that of the 3Q confinement force, 
because one cannot superpose solutions in QCD even at the classical level. 
Indeed, any 3Q system cannot be described by the superposition of 
the interaction between two quarks, 
as is suggested from the functional form (\ref{eq:Y-ansatz})
of the 3Q potential \cite{TS0102,TS0304}.

\begin{figure}[h]
\centering
\includegraphics[width=15cm,clip]{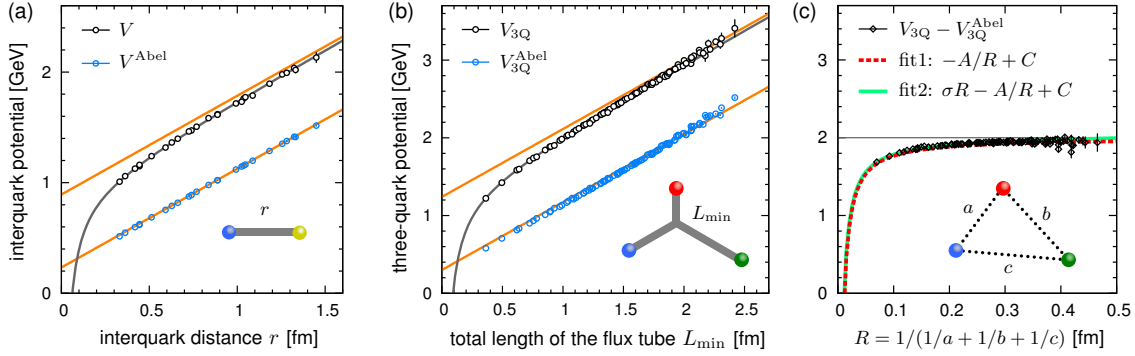}
\caption{
(a) The Q$\bar{\rm Q}$ potential $V(r)$ and 
its Abelian part $V^{\rm Abel}(r)$.
(b) The 3Q potential $V_{\rm 3Q}$ (black) 
and its Abelian part $V_{\rm 3Q}^{\rm Abel}$ (blue) 
in SU(3) lattice QCD at $\beta$=5.8 on $16^3 \times 32$.
For a rough indication, we add the best-fit Y-ansatz curve 
of the equilateral 3Q case 
for $V_{\rm 3Q}$ and $V_{\rm 3Q}^{\rm Abel}$, respectively. 
$\sigma_{\rm 3Q}$ and $\sigma_{\rm 3Q}^{\rm Abel}$ 
correspond to the slopes of the parallel lines.
(c) Fit analysis of $V_{\rm 3Q} - V_{\rm 3Q}^{\rm Abel}$.
The dashed curve is the pure Coulomb ansatz (5.3) 
with no string tension. (b) and (c) indicate 
$\sigma_{\rm 3Q} \simeq \sigma_{\rm 3Q}^{\rm Abel}$.
These figures are taken from Ref.\cite{SS15}.
}
\label{fig:main}
\end{figure}

We find that the Abelian part $V_{\rm 3Q}^{\rm Abel}$ 
of the 3Q potential also takes the Y-ansatz \cite{SS15}, 
\begin{eqnarray}
V_{\rm 3Q}^{\rm Abel}
= 
- \frac{A_{\rm 3Q}^{\rm Abel}}{R}
+\sigma_{\rm 3Q}^{\rm Abel} L_{\rm min}
+C_{\rm 3Q}^{\rm Abel},
\end{eqnarray}
with $1/R \equiv \sum_{i<j} 1/|{\bf r}_i-{\bf r}_j|$.
Figure~4(b) shows the 3Q potential 
$V_{\rm 3Q}$ and its Abelian part $V_{\rm 3Q}^{\rm Abel}$ 
plotted against the total flux-tube length, $L_{\rm min}$.
When the size of the 3Q system, $L_{\rm min}/3$, is larger than $0.3$~fm, 
$V_{\rm 3Q}$ is given by a Y-type linear potential, 
$\sigma_{\rm 3Q} L_{\rm min} +C_{\rm 3Q}$ (upper straight line in Fig.4(b)).
Remarkably, the Abelian part $V^{\rm Abel}(r)$ obeys 
$\sigma_{\rm 3Q} L_{\rm min} +C_{\rm 3Q}^{\rm Abel}$ 
(lower straight line in Fig.4(b)) at large distances, 
which means $\sigma_{\rm 3Q}^{\rm Abel} \simeq \sigma_{\rm 3Q}$.

To demonstrate 
$\sigma_{\rm 3Q}^{\rm Abel} \simeq \sigma_{\rm 3Q}$ conclusively, 
we investigate the difference $\Delta V_{\rm 3Q}$ between 
$V_{\rm 3Q}$ and $V_{\rm 3Q}^{\rm Abel}$, as shown in Fig.4(c) \cite{SS15}.
If the Abelian dominance of the 3Q potential is exact, i.e., 
$\sigma_{\rm 3Q}^{\rm Abel} = \sigma_{\rm 3Q}$, 
$\Delta V_{\rm 3Q}$ is well reproduced 
by the pure Coulomb ansatz, 
\begin{equation}
\Delta V_{\rm 3Q} \equiv V_{\rm 3Q} - V_{\rm 3Q}^{\rm Abel}
= - \frac{\Delta A_{\rm 3Q}}{R} + \Delta C_{\rm 3Q},
\label{eq:Coulomb-ansatz-DV}
\end{equation}
where $\Delta A_{\rm 3Q} \equiv A_{\rm 3Q}- A_{\rm 3Q}^{\rm Abel}$
and $\Delta C_{\rm 3Q} \equiv C_{\rm 3Q}-C_{\rm 3Q}^{\rm Abel}$.
In Fig.4(c), $\Delta V_{\rm 3Q}$ obeys 
a pure Coulomb form with no string tension, 
which is a clear evidence on the equivalence 
of $\sigma_{\rm 3Q}^{\rm Abel} = \sigma_{\rm 3Q}$, 
with accuracy within a few percent deviation, 
i.e., perfect Abelian dominance of quark confinement in baryons.

In Table~1, we summarize 
all the fit results for 
$V(r)$, $V^{\rm Abel}(r)$, $V_{\rm 3Q}$ and 
$V_{\rm 3Q}^{\rm Abel}$ 
on both lattices at $\beta=5.8$ on $16^3 \times 32$ and 
$\beta=6.0$ on $20^3 \times 32$ \cite{SS15}. 
Thus, we find perfect Abelian dominance for the string tension of 
Q$\bar{\rm Q}$ and 3Q potentials: 
$\sigma_{\rm Q \bar Q} \simeq \sigma_{\rm Q \bar Q}^{\rm Abel} 
\simeq \sigma_{\rm 3Q} \simeq \sigma_{\rm 3Q}^{\rm Abel}$.

\section{Summary and concluding remarks}

We have studied the baryonic 3Q potential in SU(3) quenched lattice QCD 
with $\beta=5.8$ on $16^3 \times 32$ and $\beta=6.0$ on $20^3 \times 32$ 
for more than 300 different patterns of 3Q systems in total, 
using 1000-2000 gauge configurations.
For all the distances, we have found that the 3Q potential is 
fairly well described by the Y-ansatz, i.e., 
one-gluon-exchange Coulomb plus Y-type linear potential \cite{SS15}. 

We have also investigated MA projection of quark confinement 
in both mesons and baryons, 
and have found perfect Abelian dominance of 
the string tension, $\sigma_{\rm Q \bar Q} \simeq 
\sigma_{\rm Q \bar Q}^{\rm Abel} \simeq \sigma_{\rm 3Q} 
\simeq \sigma_{\rm 3Q}^{\rm Abel}$, in Q$\bar{\rm Q}$ 
and 3Q potentials \cite{SS15,SS14}. 
Thus, in spite of the non-Abelian nature of QCD, quark confinement 
in hadrons is entirely and universally kept 
in the Abelian sector of QCD in the MA gauge.

\begin{acknowledgments}
H.~S. sincerely thanks Yoichiro Nambu for his interest to our studies and 
valuable suggestions in old days. 
H.~S. also thanks V.~G.~Bornyakov for his useful advices.
This work is supported in part by the Grants-in-Aid  for Scientific Research 
[15K05076, 15K17725] from Japan Society for the Promotion of Science.
The lattice calculations were done on NEC-SX8R 
at Osaka University.
\end{acknowledgments}

\end{document}